\documentclass[12pt]{JHEP3}
\usepackage{epsfig}

\title{ Geometrical quantities of  fuzzy sphere}
\author{Jingbo Wang \\
  Non-equilibrium Condensed Matter and Quantum Engineering Laboratory, Department of Applied Physics,
 Xi'an Jiaotong University,
 Xi'an, 710049, People's Republic of China\\
 \email{ shuijing31@gmail.com }}
\author{ Yanshen Wang \\ Non-equilibrium Condensed Matter and Quantum Engineering Laboratory, Department of Applied Physics,
 Xi'an Jiaotong University,
 Xi'an, 710049, People's Republic of China\\
 \email{  yswang@mail.xjtu.edu.cn }}
\abstract { In this paper, we consider the geometrical quantities on
the fuzzy sphere from the spectral point of view, such as the area
and the dimension. We find that, in contract to the standard sphere,
the area and the dimension are the functions of the energy scale of
the fuzzy sphere.}

 \preprint{} \dedicated{} \maketitle
\begin{document}
\section{Introduction}
Fuzzy space \cite{1} is a kind of noncommutative space, and it can
provides an alternative way of regularizing field theories with
finite degrees of freedom \cite{2}. A good review of field theory on
fuzzy sphere is \cite{3}. Recently the non-perturbative numerical
studies of quantum fields on fuzzy sphere attracted many attention.
It has been shown that there exist three distinct phase in the
$\phi^4$ real scalar field theory \cite{4,5}. Also the phase
structure of Yang-Mills-Chern-Simons field was given in \cite{6}.

In paper \cite{7}, we consider the behavior of the fuzzy sphere
$S_F^2$ with the help of the spectral action, and find that in
different energy scale, the fuzzy sphere behaves differently. In
this paper, we consider two important geometrical quantities, that
is the area $A$ and the dimension ${D_s}$, from the spectral point
of view. We find that, unlike to the standard sphere, those two
quantities are functions of the energy scale $\Lambda$.

The paper is organized as follows. In section two we give the
spectrum of the Dirac operator on the fuzzy sphere. In section three
we calculate the area of the fuzzy sphere. In section four we
calculate the dimension of the fuzzy sphere. In section five we
prevent the conclusion and the discussion.
\section{The spectrum of the Dirac operator}
The algebra $A_N$ of fuzzy sphere $S_F^2$ is generated by the
operators $x_i$ (i=1,2,3) satisfying
\begin{equation}
[x_i ,x_j ] = i\frac{{2l}} {{\sqrt {N(N + 2)} }}\varepsilon _{ijk}
x_k ,\label{eq1}
\end{equation}
with constrain
\begin{equation} x_i x_i = l^2 .
\label{eq2}
\end{equation}
Each operator of $A_N$ is represented by a matrix acting on the
$N+1$-dimension Hilbert space $F_N$. Apparently, the matrices $x_i$
can identified with the generators of $su(2)$ Lie algebra and the
generated algebra is equivalent to the algebra of $(N+1)\times
(N+1)$ matrices, $ M_C (N + 1) $.

Dirac operator can be defined on the fuzzy sphere \cite{8}. We cite
the spectrum of the Dirac operator in \cite{8}. The spectrum of the
Dirac operator $D$ on the fuzzy sphere with $l=1$ is as follows,
\begin{equation} D^2
\Psi _{jm} = \lambda _j^2 \Psi _{jm} ,\label{eq3}
\end{equation}
 with
\begin{equation}\lambda _j^2  = (j +
1/2)^2 (1 + \frac{{1 - (j + 1/2)^2 }}{{N(N + 2)}}) , \label{eq4}
\end{equation}
where $j$ and $m$ are half integer and take the value \[ 1/2 \le j
\le N + 1/2, \qquad - j \le m \le j .
\]From the expression \ref{eq4} we can see that, unlike to the standard sphere, those eigenvalues are finite and not
monotonically increasing.
\section{The area function}
To calculate the area of the fuzzy sphere, we have to define it
though the spectrum of the Dirac operator. On standard sphere, we
have the heat kernel expansion for the Dirac operator \cite{9}.
\begin{equation}
Trace(e^{ - tD^2 } ) = \sum\limits_{n = 1}^\infty  {4ne^{ - n^2 t} }
\simeq 1/(2\pi t)A - 1/3  +  o(t) ,\label{eq7}\end{equation} as $t
\to 0$, where $A$ is the area of the sphere. So we have, for the
standard sphere,
\begin{equation}
A \approx 2\pi t(Trace(\exp ( - tD^2 )) + 1/3),
\label{eq7}\end{equation} if $t$ is not too big. Then we generalize
this formula to the fuzzy sphere, that is, we define the area of the
fuzzy sphere as follows,
\begin{equation}
A \equiv 2\pi (Trace(\exp ( - D^2/\Lambda^2 )) + 1/3)/\Lambda^2,
\label{eq7}\end{equation} where $t=1/\Lambda^2$, so $\Lambda$ has
the dimension of energy. Then with the spectrum given by \ref{eq4},
we can calculate the area of the fuzzy sphere for every integer $N$.
\begin{equation}
{A(\Lambda ) = 2\pi (\sum\limits_{l = 1}^N {(4l)} \exp (-\frac{{l^2
}}{{\Lambda ^2 }}(1 + \frac{{1 - l^2 }}{{N(N + 2)}})) + 1/3)/\Lambda
^2 }. \label{eq7}\end{equation} Since we don't have an analytic
expansion for the area function, we just drawn it out. See figure 1.

\EPSFIGURE{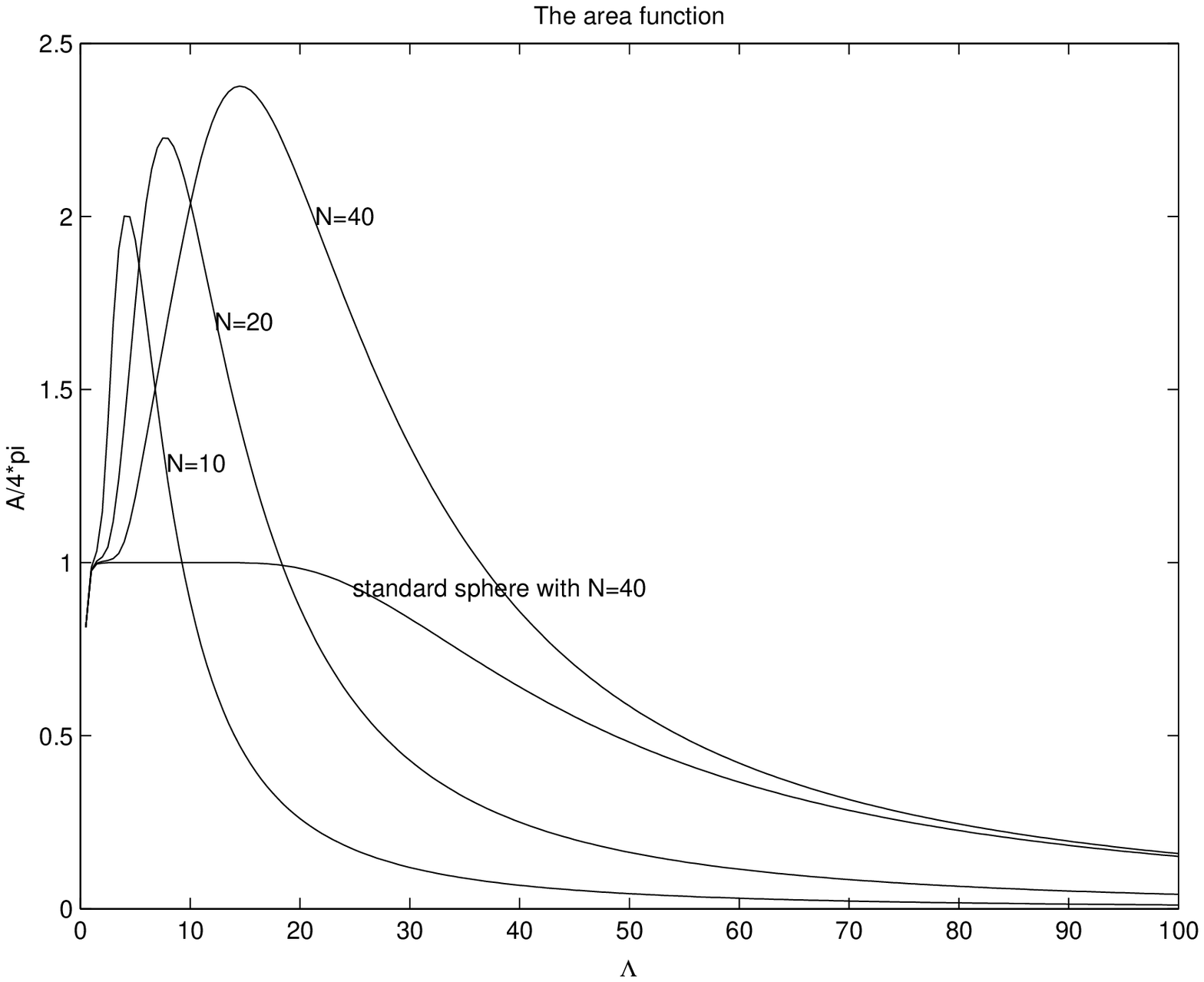} {The area function for standard sphere and fuzzy
sphere with different $N$}

In the range $(0,1)$, since the heat kernel expansion breaks down,
we don't rely on this figure. In this range, the first few
eigenvalues play important role, so the fuzzy sphere behaves like
the standard sphere. See \cite{7} for the more discussion of this
problem. From this figure we can see that for the standard sphere
the area function is not constant. The reason is that the number of
eigenvalue of the Dirac operator for the standard sphere is
infinite, but we just choose the first $40$ terms. The more terms we
choose, the more closed the area function to constant $1$. The area
function for the fuzzy sphere is not constantly at all. The area
function achieves a max value and then decrease to zero at high
energy. With $N$ increasing, the max value of the area function
increases too.
 \section{The fractal dimension}
In the paper \cite{7}, we find that, from spectral point of view,
the fuzzy sphere can have different dimension for different energy
scale. Since this result is very rough, we give more investigation
to this issue. Fractal dimension for quantum sphere \cite{10} and
for loop quantum gravity \cite{11} has been discussed, and we borrow
the operational definition from those two papers, that is, the
dimension for the fuzzy sphere is,
\begin{equation}
D_s  =  - 2\frac{{d\ln P(T)}}{{d\ln T}}, \label{eq7}\end{equation}
In our case, $P(T)=Trace(-D^2/\Lambda^2)$ and $T=1/\Lambda^2$, so
\begin{equation}
D_s  =  - \frac{{d\ln (Trace(e^{ - D^2 /\Lambda ^2 } ))}}{{d\ln
\Lambda }}. \label{eq7}\end{equation} Let's just figure it out for
the fuzzy sphere. See figure 2.

\EPSFIGURE{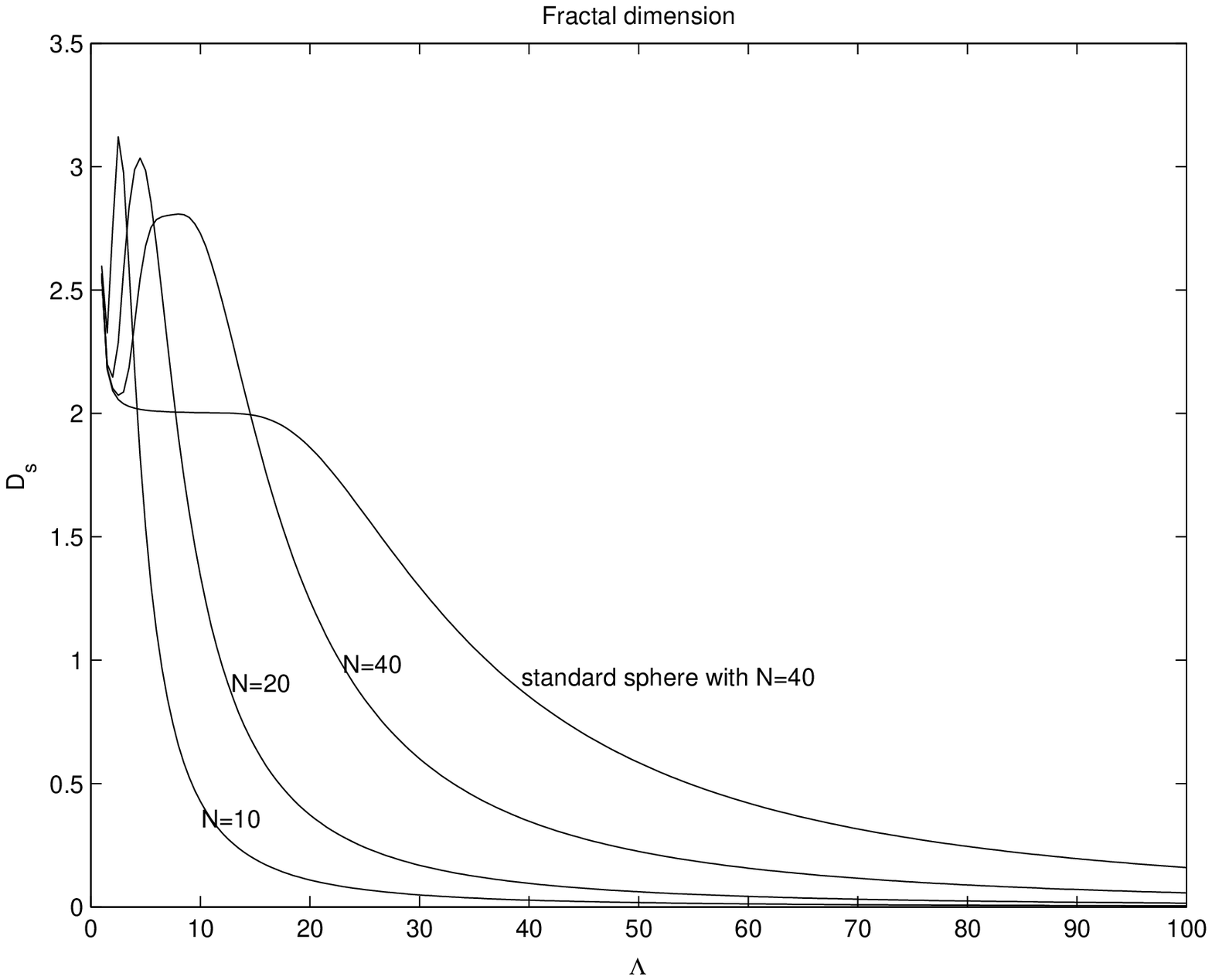} {The fractal dimension for standard sphere and
fuzzy sphere with different $N$}

This figure looks like the first figure. The dimension of the
standard sphere is not constantly $2$ because we just choose the
first $40$ terms. The fractal dimension for the fuzzy sphere changes
when the energy scale $\Lambda$ changes, and decay to zero at high
energy. When $N$ becomes large, the max value of the fractal
dimension becomes small and approach the standard dimension $2$.
\section{Conclusion}
Dirac operator play an important role in noncommutative geometry
\cite{12}. In paper \cite{13}, the authors show that the eigenvalues
of the Dirac operator can be treated as the dynamical variables for
the Euclidean general relativity. From this point of view, every
geometrical quantities must be function of those eigenvalues. In
this paper, we calculate the area and the dimension of the fuzzy
sphere from the spectrum of the Dirac operator. We find that they
are functions of the energy $\Lambda$. We hope that the operational
formula for the area and the dimension of a manifold will play some
role in quantum gravity. \acknowledgments{ The authors have been
supported by the National Natural Foundation of China (grant
no.10375045)}


\begin{thebibliography}{99}
\bibitem{1}J.Madore, \cqg{9}{1992}{69}
\bibitem{2}H.Grosse, A.Strohmaier,\lmp{48}{1999}{163}
\bibitem{3}A.P.Balachandran, S.Kurkcuoglu, and S.Vaidya,  {\it Lectue on fuzzy and fuzzy SUSY physics }
\hepth{0511114}
\bibitem{4}M.Panero, \jhep{0705}{2007}{082}
\bibitem{5}D.O'Connor and C.Saemann, \jhep{0708}{2008}
{066}
\bibitem{6} T.Azuma, S.Bal, K.Nagao, and J.Nashimura, \jhep
{5}{2004} {005}
\bibitem{7} Jingbo Wang and Yanshen Wang, \cqg{26}{2009}{ 155008}
\bibitem{8}U.Carow-Watamura, S.Watamura, {\it Differential
calculus on fuzzy sphere and scalar field}, \ijmpa{13}{1998}{3235}
\bibitem{9}P.Gilkey, {\it Invariance theory,the heat equation and
the Atiyah-Singer index theorem,} 2nd editon,Boca Raton:CRC Press
1995
\bibitem{10}D.Benedetti, \prl{102}{2009}{
111303}
\bibitem{11} L.Modesto, {\it Fractal structure of loop quantum
gravity},\hepth{0812.2214}
\bibitem{12} A.Connes, {\it Noncommutative Geometry,} Academic
Press, New York, 1994
\bibitem{13} G.Landi and C.Rovelli, \prl{78}{1997}{
3051}
\end{thebibliography}
\end{document}